\documentclass[12pt,halfline,a4paper]{ouparticle}

\usepackage{url}

\begin{document}

\title{The gift of the gab:\\
Are rental scammers skilled at the art of persuasion?}

\author{%
\name{Sophie Van Der Zee, Richard Clayton, Ross Anderson}
\address{Computer Laboratory, University of Cambridge,\\JJ Thomson Ave, Cambridge, CB3 0FD, UK,}
\email{Sophie.Van-Der-Zee@cl.cam.ac.uk\\
Richard.Clayton@cl.cam.ac.uk\\
Ross.Anderson@cl.cam.ac.uk}
}

\abstract{
Rental scams are a type of advance fee fraud, in which the scammer tries to get a victim to pay a deposit to rent an apartment of which the scammer pretends to be the landlord. We specifically focused on fraudulent long-term rentals advertised in the UK on Craigslist. After a victim responds to the scammer's advertisement, the scammer attempts to persuade them to transfer money without having seen the property. We were interested in which persuasion techniques scammers use, and in assessing their skill at the art of persuasion. During a period of three weeks, we scraped 2\,112 letting advertisements, identified the fraudulent advertisements and had 44 conversations of around 4 or 5 emails each with the scammers. Our analysis indicates that Cialdini's marketing-based social persuasion strategies, such as liking, appeal to authority, and the need for commitment and consistency are extensively implemented by rental scammers. Of Stajano and Wilson's scam-based persuasion strategies, an appeal to sympathy (i.e., kindness) and need for greed were commonly used. We identified two further social persuasion strategies: establishing credibility and removing objections. At a superficial level, rental scammers seem skilled at their job, because they mimic genuine landlords and use a range of effective persuasion techniques. However, when examining their emails more closely, we see they often use pre-scripted emails, their mimicry is often incompetent, and they have a lack of language skills and cultural knowledge that may tip people off. They appear to be the criminal equivalent of a boilerhouse sales operation, a modus operandi that has not previously been studied by cybercrime researchers. 
}

\date{\today}


\maketitle

\section{Introduction}
\label{sec1}

Fraud, the act of deception for personal gain, and its deterrence, can be traced back to ancient times. It is a recurring theme in Greek mythology, with many stories about Dolos (the spirit of trickery and guile) and Apate (the goddess of fraud and deceit). Although fraud has occurred throughout history, its prevalence and shape have changed over time. Technological developments such as the post, telephone and fax enabled fraudsters to target potential victims more efficiently, removing the need for face-to-face contact~\cite{Zingerle}. The latest evolution started in the nineties when the use of email became widespread~\cite{Buchanan, Holt}; as more and more people went online, scammers started to use the Internet to target millions of people worldwide while concealing their true identity and location~\cite{Edelson}.

Although the Internet has affected a range of existing fraud types and created several new ones, it has greatly increased the ease and incidence of advance fee frauds (AFF). Also called `419 scams' from the relevant article in the Nigerian Criminal Code, these typically involve tricking a potential victim into transferring money by persuading them that they will shortly receive a larger sum of money in return~\cite{Chang}.

Scammers experiment with different story-scripts~\cite{Zingerle}, but a classic tactic is to impersonate a wealthy and powerful person with access to a large sum of money~\cite{Edelson}. The scammer's tale is that before this fortune can be accessed, he needs an accomplice; then a small payment needs to be made, for example to cover processing fees, unforeseen taxes or insurance. The collection of these `advance fees' is the real objective of the scam~\cite{Onyebadi}.

AFF in its current international form, originated in Nigeria in the seventies and is still largely driven by fraudsters from West Africa~\cite{Chang, Onyebadi}. A sense of historical revenge provides a sense of justification to at least some scammers; during the trial of Nigerian lawyer and alleged fraud kingpin Fred Ajudua, for a series of 419 offences, he claimed that the frauds were ``compensation from white men for slavery and colonialism''~\cite{Glickman, NewsNigeria}.

\subsection{Rental scams}

A specific type of AFF is the rental scam. We came across different types of rental scams, including short-term holiday rentals and long-term letting. On the long-term rental market, scams can be divided between those where an apartment can be viewed before payment is made, and scams in which viewing is not possible~\cite{Independent}. In the latter, the victim responds to an advert for a property and pays money up front without viewing the place. The scammer's advertisement may make the property seem especially desirable or the rent may be set below the market rate. Once a potential victim answers, emails are exchanged and then a reason is found for the victim to transfer money. 

Although not being able to view a property before renting would serve as a red flag for some people, it may not put off students and other people who move across country or even across borders, as they may have no choice but to rent sight unseen. In other cases, the scammer is prepared to offer a viewing but only once a fee has been paid to cover a credit check, postage for sending keys for a viewing, or a deposit may be requested along with advance rent. In some cases the victim may just be asked to make a Western Union payment to a friend to show that they have sufficient funds, but the scammer impersonates the friend and steals the money.

Although rental scams have not previously been studied in the literature, they are widespread. Based on an online survey by Shelter in 2010, the UK National Fraud Authority estimated that all rental scams together annually cause 315,000 victims to lose on average of \pounds 2,394, leading to an overall cost of \pounds 755 million a year~\cite{NFA}. However, the Shelter survey covered many other types of rental scams besides the AFF version that we discuss here~\cite{Shelter} and in particular it included disputes where a landlord refuses to refund a tenant's deposit. 

As an order of magnitude estimate for the AFF component of the overall fraud figure, we attempted to determine how many Cambridge students had been affected by such frauds in 2014. The police received 44 victim reports, but all appear to be for a specific scam that targeted students with fake websites, local phone numbers and some convincing paperwork. Scaling this up across all UK students (of which around 1\% are in Cambridge) would imply an overall impact of around \pounds 4 million per annum, but this will be on the high side because the scam was specific to Cambridge and a handful of other towns, and was rather more convincingly executed than the scams that we look at in this paper. Equally not all victims are students and not all victims report their loss to the police -- so our best estimate is that the annual take from AFF style rental scams in the UK is in the range \pounds 2--10 million -- and may be of a similar magnitude in several other countries because we have also seen scams targeting rentals in other European and US cities.

We personally got interested in rental scams when the first author encountered a fake letting advertisement when moving to Cambridge for a post-doctoral research position. Fortunately, she realised it was a scam before transferring money, but it made her realise how well-crafted this type of AFF could be. Since then, she has been in contact with several people at the University of Cambridge who have lost money this way. The experiences of four scam victims are described later in this paper.

Other indications that rental scams are becoming a serious issue is provided by large housing websites such as Craigslist, Trulia, and Zillow, which now explicitly mention rental scams in their help section on ``how to avoid getting scammed''~\cite{Craigslist, Trulia, Zillow}. Although many fraudsters appear to operate from outside the jurisdiction, in some cases criminals have been caught and prosecuted.

\subsection{Why do scams work?}

To understand why scams succeed, and to prevent others from falling victim in the future, researchers have tried to determine why victims fall for scams. For example, Modic and Lea investigated how personality traits correlate with the likelihood of being scammed, and found that traits such as extraversion and openness can put one at risk~\cite{Modic}. Although an interesting result, it is not obvious how we might use it to reduce the number of incidents.

An alternative approach involves studying the methodology of scams~\cite{Whitty}. Previous research has demonstrated that scammers tend to borrow social persuasion strategies from the marketing industry~\cite{Stajano, Cialdini}. For example, interviews with real victims show that scammers frequently appeal to scarcity (highlighting a limited offer), authority (forging emails from the FBI) and exploit people's greed (stressing the amounts of money involved). Carter's text analysis of postal scams revealed that scammers appeal to scarcity, but also inspire secrecy~\cite{Carter}.

Wilson experimented on unwitting members of the public using a variety of (staged) scams in the BBC TV program ``The Real Hustle''~\cite{Stajano}. Stajano and Wilson demonstrated that these scams implement all six social persuasion strategies identified by Cialdini: reciprocity, commitment and consistency, liking, authority, scarcity and social proof~\cite{Cialdini}, plus the `need and greed' principle described by Lea et al.~\cite{Lea}. Their analysis also identified three additional persuasion techniques: `distraction', `dishonesty' and `kindness', leading to a total of ten scam-related persuasion techniques. However, their principle of `kindness' expressed as ``people are fundamentally nice and willing to help'', typically revolves around the telling of sad stories, and so we have renamed it to the more intuitive notion of appealing to `sympathy'. 

The majority of research on AFF has focused on interviews with victims but some researchers have analysed the initial letter or email sent by the scammer~\cite{Onyebadi, Carter}. However, most AFF does not involve a single email. Instead, the scammer and victim often exchange several emails before money is transferred and the scam is completed. From the interactive communication literature we know that interaction partners influence each other, so the behaviour of one person depends on the behaviour of the other~\cite{Garrod}. In this paper, we analyse how the interaction between scammer and victim develops over time by analysing email exchanges. This analysis also provides an opportunity to study whether the criminals appear to be skilled in the arts of persuasion.

\subsection{Current study}

The current study provides us with some basic statistical data on the prevalence of rental scams on Craigslist along with some indications as to the number of criminals involved. We are also able to explain the basic mechanics of the fraud, how the criminals construct the fake advertisements and the types of stories they tell to justify an advance payment.

We then discuss the types of persuasion techniques the criminals use. Their methods have been coded based on Cialdini's social persuasion categorization, including reciprocity, commitment and consistency, liking, authority, scarcity and social proof~\cite{Cialdini}. Each email was also coded for the additional scam principles `need and greed'~\cite{Lea}, `distraction', `dishonesty' and `sympathy'~\cite{Stajano}. We then found that there were further relevant techniques which had not been described by these authors; we dub them `establishing credibility' and `removing objections'.

We assess the criminals' competence at using these methods, and also whether they were sticking to a pre-determined script or were capable of adapting to circumstances as the email conversation progressed. In other words, we consider not just the scenarios that might convince the victim to send money, but also the way in which the criminal tries to persuade the victim that they should act. 

To answer these research questions, we gathered a set of fake letting advertisements and subsequently interacted with the advertisers through email. Because our interactions with the scammers stopped before an actual money transfer had taken place, we complete the picture by including the self-reported experiences of four real victims of rental scams.

\section{Methods}
\label{sec2}

\subsection{Procedure}

In early 2014, we ran an experiment in which, for three consecutive weeks, we scraped letting advertisements off the classified advertising website Craigslist across 27 UK cities, after which we sent email replies to a large number of apparently fake adverts. Our aim was to gain a better understanding of rental scams and the methods scammers use to persuade victims to transfer money. 

During the experiment we manually categorized each advertisement as `genuine', `fake', `unsure' or, when the advert had been placed in the wrong category, 'irrelevant'. When an advertisement was categorized as fake, or when we were unsure, an automated email response was sent to the advertiser.

To make it harder for scammers to recognize the true source of our emails, we created a webmail account for each advert with a different `person' operating each one. We constructed the responders' unique names by randomly combining common British first and last names. All the first names were unisex (e.g., Ashley, Rowan, Dylan, and Charlie). These names were checked for plausibility by three British English native speakers, and only included when all three agreed. This way, the gender of the victim was initially unknown to the scammer.
    
Each of our initial replies to the adverts was automatically created by randomly selecting alternatives from a pre-set framework, set out in Figure~\ref{fig_lettertext}. Although the word use and content of each automated email was unique, our messages always displayed an interest in the advertised property, provided some personal information, asked a question about the property and included a request for a viewing.

\begin{figure}[!t]
\caption{ Framework for the first paragraph of the response email showing the alternative phrasing for each part of the sentences that were constructed.}

\begin{quote}
\begin{texttt}
\{I am\textbar I'm\} \{very\textbar very much\textbar really\textbar extremely\textbar genuinely\} interested in \{your advertisement\textbar your property\textbar your house\textbar your place\textbar the property you advertised\textbar the place you advertised\textbar the house you advertised\}. It \{seems \textbar looks \textbar sounds\} \{great! \textbar amazing! \textbar nice. \textbar really nice. \textbar lovely. \textbar like a good opportunity. \textbar a good deal.\}
\end{texttt}
\end{quote}
\label{fig_lettertext}
\end{figure}
On receiving an answer from the advertiser, the original advertisement was re-evaluated. When an advertisement now seemed genuine, we sent an apology email, explaining we were no longer interested in the property. When an advertisement was still categorized as unsure, we continued the interaction with the advertiser. For all advertisements categorized as fake, we started an email conversation with the advertiser to learn about the scam process and the social persuasion techniques used. For ease of interpretation, the advertiser will henceforth be called the ``scammer'', and the experimenter the ``victim''. 
    
After the initial template-based email, all further email contact was conducted manually following a set of guidelines. First, all chosen names were purposefully gender neutral, but upon request the victim revealed she was female. Second, to indicate employment and income security, the victim would state that they had just been offered a job with a 1/2/3 year contract at the local university in a variety of disciplines. Third, when asked, the victim mentioned she had to move from one UK city to another for her new job, implying that she was not familiar with the housing on offer. Fourth, to minimize the chance of being rejected, the victim did not smoke and did not have pets.

In the follow-up emails, the researcher manually responded to the scammer's questions if applicable, while displaying a continuous interest in the property and repeatedly requesting a viewing. This strategy was repeated until the scammer first requested a money transfer. Then a three step procedure was followed. In the first two emails after the money transfer request, the victim replied showing hesitation about the transfer but a continued interest in the property. These emails usually included a proposed alternative to the money transfer (e.g., a deposit and first month's rent payment in cash upon satisfactory viewing and contract signing), or a request for more information. In the third email after the financial request the potential customer made clear that she would not transfer money without viewing the property, and that she was no longer interested in the property. This procedure gave scammers a maximum of three emails to convince the victim to transfer the money. In two cases (ID~75 and 78) we accidentally stopped responding but the other conversations ran until the protocol completed or the scammer failed to reply. 

\subsection{Dataset}

Over a period of three weeks, from 12 February to 3 March 2014, we automatically scraped 2\,112 advertisements that appeared on the UK accommodation sections of Craigslist. We categorized each advertisement based on its content, language use and the characteristics of the accommodation using four simple heuristics:
    
\begin{enumerate}

\item
When something seemed too good to be true, we assumed it was a scam. For example, low rental prices or pictures of amazing interiors were a reliable indicator.

\item
We deemed it a scam when text seemed more appropriate for holiday rentals, such as the provision of fresh towels on a daily basis. Misplaced adverts for holiday lets were all ignored, whether or not they appeared to be a scam, because we only wished to consider adverts for longer periods of rental.

\item
Scams often misused capital letters or had grammar and spelling mistakes. Again this was only an indicator because this heuristic also applied to some genuine individuals, predominantly those renting out student rooms.

\item
The most reliable criterion involved googling phrases and pictures from the adverts to see if they appeared anywhere else. We found that text and pictures were often copied verbatim from a genuine website, with the only difference being a lower price, and often a different location. For example, luxury apartments in Manhattan, US, were suddenly for rent in London for a quarter of the price.
\end{enumerate}

Using these guidelines, 119 adverts were classified as either `fake' or `unsure'. After email interaction with each advertiser, we decided that 55 adverts were truly scams, 29 were genuinely posted by an individual, 23 were genuinely posted by a legitimate business and we remained unsure about the remaining 12. Of the 55 scam conversations, we removed an additional 11 because two involved immediate redirection to a dubious website so there was no email conversation. The other nine turned out to be for short-term holiday accommodation -- different rental scams that fall outside the scope of this paper. This left us with a dataset of 44 scam conversations.

In addition to analysing the data per email or per scam conversation, we also attempted to take into account that some conversations may have been held with the same person. We tried to identify them based on their name, email address and copy-pasted email content. We believe that we were communicating with 21 different scammers, with whom we had between one and six scam conversations each (M = 2.10, SD = 1.81). However, this is only an estimate and the true number may have been less.

Our email interactions stopped before any money was transferred, because that would involve crossing an ethical and possibly a legal boundary. However, to be able to understand what happened in rental scams after the money transfer we invited real rental scam victims to fill out a questionnaire about their experience, which four people did. Although this is too few for statistical analyses, the information does provide insights into the rental scam process from the victim's point of view. The questionnaire consisted of demographic questions such as gender (50\% female), age (18--25), education (all BSc or higher), and profession (all students), and included questions about their scam experience. We asked about where they saw the advertisement, the pretext for the payment, how much money was lost and how they found out they had been scammed. Respondents were also asked to indicate the financial and emotional impact the scam had on them. The answers are discussed in the results section. 

To summarise our dataset: from inspecting three weeks of letting advertisements on Craigslist we held 44 email conversations with scammers and we added to this the information provided by four real victims of rental scams. This rich and interactive dataset has enabled us to identify scam characteristics, to learn how scammers persuade their potential victims to transfer money, and ultimately, to determine if the scammers are skilled at the art of persuasion. 

\subsection{Coding persuasion data}

Our dataset not only includes the initial advertisement, but also the emails in which the scammer attempts to persuade the victim to transfer money. To establish if scammers are any good at the art of persuasion, we manually coded each scam email for established social persuasion techniques.

We initially coded a total of ten persuasion techniques, including Cialdini's reciprocity, commitment and consistency, liking, authority, scarcity and social proof~\cite{Cialdini}, Lea et al.'s need and greed~\cite{Lea}, and Stajano and Wilson's distraction, dishonesty, and sympathy~\cite{Stajano}. While Cialdini's six persuasion techniques were based on human behaviour in general, the remaining four techniques were identified by analysing actual scams.

In addition, for each email we analysed whether the scammer made spelling and grammar mistakes. Abbreviations and `texting' language such as ``pls'' for ``please'' and ``u'' for ``you'' were not counted as spelling/grammar mistakes, as they may have been intentional. Because we were trying to measure lack of English skills and not sloppiness, we overlooked the first mistake in each email, but any further errors meant that the email was deemed to contain `mistakes'.

We now describe the ten persuasion techniques that we were looking for, discussing each of them in the context of rental scams.
 
\subsubsection{Reciprocity (Cialdini,~\cite{Cialdini})}

Reciprocity is the human tendency to return what we get; if someone does us a favour, we are more likely to do a favour in return.  Scammers can use reciprocity by offering something small first, in order to increase the chance of the victim transferring the funds. Alternatively, the scammer can copy the victim's mood, for example by showing hesitation or even aggression when the victim starts doubting the legitimacy of the deal (see ID~92). 

\begin{quote}
ID~92. \textit{As I told you UPS had this service and I have used it in the past. It is simple. They will hold your money until you visit the apartment and if you are not satisfied they will transfer the amount back into your account. You won't lose anything. So please think again and if you agree just give me your full name and address so we can get started.}
\end{quote}

\subsubsection{Commitment and consistency (Cialdini,~\cite{Cialdini})}

This principle, also known as the `foot in the door' technique, encompasses people's wish to continue something they have already started so as to be consistent. For example, the scammer can show his commitment by claiming he has reserved the apartment `especially for you'. He can then ask for a token of commitment, by asking the victim to transfer a deposit to show they are serious about renting the apartment, as demonstrated by ID~10:

\begin{quote}
ID~10: \textit{The 100 is not postage fee, is part of deposit to reserve the place for you because other tenant also making enquiry about the place. If you are comfortable with that, I will mail the keys to the building manager.}
\end{quote}

\subsubsection{Liking (Cialdini,~\cite{Cialdini})}

People are more likely to be influenced by people they like. A scammer can increase the chance of being liked by being friendly, caring, giving compliments, or even emphasizing similarities with the victim. For example, the scammer in ID~31 combined being friendly with listing his traits, habits and preferences in order to increase the victim's chances of finding similarities:

\begin{quote}
ID~31: \textit{About Me: i am a simple person. simple and uncomplicated. i enjoy the simple joys and pleasures that life has to offer. like coffee when i wake up in the morning or whenever i feel like having a cup. listening to music reading or watching a movie. just being able to wake up in the morning. laughing spending time with friends ( i don't have that many) spending time with flatmates and best of all spending time with my boyfriend. i don't like things that are complicated or too complex. i like things that flow together easily seamlessly. i want things that are cut and dried things that make sense. i like answers like yes or no, i don't like maybe because it is so uncertain. clear answer make me feel a certain sense of security. and that feeling of security is vital to me. it's like water or food for me, i like meeting new people who are ready to share experience, talk and a cup of coffee and watch movie with me as well.}
\end{quote}

ID~13 explicitly highlighted the similarities between victim and scammer:

\begin{quote}
ID~13: \textit{I'm happy that you are a woman like me.I'm with my mum now and I can only take the next available flight to england if you can send a month money to any of your trusted friend in UK via money gramm transfer or western union money transfer.}
\end{quote}

\subsubsection{Authority (Cialdini,~\cite{Cialdini})}

An appeal to authority relies on people's sense of obligation and respect to people in authority, and as a result, people are more likely to be influenced by an authority figure than by people at random. Authority can be induced in different ways, including wearing a uniform, displaying a title or important job, and having accessories such as expensive gadgets and cars. Rental scammers frequently refer to their lawyer or solicitor in order to give their proposal extra weight, as was done by ID~2. Alternatively, scammers can try to increase their influence by mentioning their own well-paid job or important title. 

\begin{quote}
ID~2: \textit{Thanks for getting back to me. I wouldn't want you to get me wrong, the idea of making a confirmation of your financial ability is not only my idea, my solicitor and i decided to use that process and i can't violate my solicitor procedure. I am a God fearing woman and i am not after your money. I have my own work and i earn enough money for my family and also to help the less privileged.}
\end{quote}

\subsubsection{Scarcity (Cialdini,~\cite{Cialdini})}

Limiting the availability of an offer will increase the demand. Availability can be limited through time pressure or by limiting the stock. In general, scammers emphasised time pressure with statements such as `as soon as possible', `now' and `quickly'. Rental scammers also demonstrated scarcity by mentioning that other people were interested in the apartment as well, pushing the victim to decide quickly. For example, ID~36 forwarded an email from another potential tenant, demonstrating that the apartment was in demand. 

\begin{quote}
ID~36. \textit{Hello Natalie, Good afternoon hope your doing great? Just so you know I have stopped responding to other people interested in the flat here is another info of a tenant wanting the flat urgently,}

Email of other potential tenant: \textit{My full name is XXXXX, I'm new to New Hampshire from Miami Florida and I'm looking for a place where me my daughter and my fianc\'{e} can live. My fianc\'{e} can provide all occupation information as I am currently seeking for employment, but his income alone can cover the monthly cost. He is the manager at the XXXXX restaurant in XXXXX. We have no pets. My expected move in date is march 15th. My move out date depends on when you expect for me to move out. As long as I can rent at least a year I am fine with everything, But be rest assured the flat is yours sincerely to me you sound more serious i would be waiting to hear from you as soon as possible.}
\end{quote}

\subsubsection{Social proof (Cialdini,~\cite{Cialdini})}

People are more likely to do something if they have proof that other people are doing the same thing. Especially in uncertain or new circumstances, people tend to look at others before deciding what to do. Scammers can use this need for social proof to their advantage, by explicitly mentioning their previous renters went through the same procedure and were happy with the outcome (as ID~26 demonstrates). More deceptively, the scammer can offer the contact details of a `happy previous tenant', so they can be directly asked about their experiences.

\begin{quote}
ID~26. \textit{I would like you to know that i am not after your money as i have my own work and i know what it takes to earn a living and also i thought i have explain the whole process to your understanding as you need to make a payment before we can proceed with the viewing as the tenant that left my flat from Oxford also follow the same process and there wasn't any problem.}
\end{quote}

\subsubsection{Need and greed (Lea et al.,~\cite{Lea})}

Wanting or even needing something creates vulnerability because it will cause people to be more easily persuaded. A rental scammer can appeal to a person's need and greed in two ways: first, by offering something the potential victim is likely to want, such as a great apartment for low cost, as is exemplified by ID~16; secondly, they can emphasise how great a deal, or `once in a lifetime opportunity' their apartment is.

\begin{quote}
ID~16: \textit{This is one of the best apartments you can get at this reasonable rate. The apartment is available for your dates but I cannot guarantee the hold of the apartment until booking is made.}
\end{quote}

We created an additional need and greed measure for each complete interaction based on rental price and location, using a set of threshold values (London, \pounds 400 for a room and \pounds 800 for an apartment; South England, \pounds 350 and \pounds 700; North England, Scotland and Ireland, \pounds 200 and \pounds 400). When the advertised rent was below the threshold, the interaction was coded as an appeal to need and greed. 

\subsubsection{Distraction (Stajano and Wilson,~\cite{Stajano})}

When distracted or even blinded by something of interest, people are less likely to think of possible threats. For example, a scammer may try to distract their victim by pointing out further benefits of their apartment, though most of the examples in our dataset also invoked sympathy, such as the request for a loan to assist the sick mother of ID~13:

\begin{quote}
ID~13: \textit{Oh miss Izzel,you really diapoint me.there are many people wanted to get the appartment and I told them it has been occupied by you.I told you my mum is sick and she's in the hospital. Anyway goodluck with your new appartment but I will apriciate if you can borrow me money to pay the bills here in the hospital and buy some drugs for here.please I'm waiting for your response.I promise I will return it when she get back and I found someone that will get my property.hope to hear from you soon.thanks}
\end{quote}

\subsubsection{Dishonesty (Stajano and Wilson,~\cite{Stajano})}

Some scams cause the victim to engage in illegal behaviour, which can be used to persuade them not to report the scammer. For example, in a classic 419 scam, the victim is asked to impersonate a relative of a recently deceased rich man so as to receive an inheritance from a crooked executor. However victims will not commit offences when renting property and we found no scammers using this persuasion method.

\subsubsection{Sympathy (called ``kindness'' by Stajano and Wilson,~\cite{Stajano})}

People tend to help others, so appealing to someone's sympathy can influence their behaviour. Rental scammers can emphasize their difficult situation and ask you to help them. For example, they may mention an ill mother they must take care of, or that they have had a bad experience with other renters in the past.

\begin{quote}
ID~58. \textit{Thank you for the response to my listing, I'm the owner of the apartment you are making inquiry of. Actually my parents resided in the apartment before i lost both parents in a ghastly motor accident\footnote{A ``ghastly accident'' is somewhat of a clich\'{e} in 419 scam emails -- the term being somewhat dated and jarring to modern British ears.} so i had to relocate to Africa to with my fathers half brother running a charity home in Africa.(W.AFRICA) and presently my apartment is still available which i inherited from my late parents for rent.You where actually supposed to be sharing apartment with me but since i had to relocate i need someone who can look after the apartment for now.}
\end{quote}

\section{Results}
\label{sec3}

Our scam conversation dataset of 44 email conversations, of four to five emails each, plus the scraped advertisements and the gathered victim experiences, was analysed to provide an insight into a number of different topics. We wanted to be able to describe the key characteristics of rental scams; we wished to understand which social persuasion strategies rental scammers use; and we wanted to determine how good the scammers were at their task. We will conclude this section with a discussion of what we learnt from four real scam victims.

\subsection{Scam characteristics}

\subsubsection{Origin}

Advance fee fraud is generally believed to be a speciality of criminals from West Africa~\cite{Park}. In this study, most of our correspondents took advantage of Craiglist's email relay system which hides the true identity of our correspondents, so we cannot form a view as to whether rental scammers fit into this general pattern.

We can get a small hint by analysing the telephone numbers that six (14\%) of the scammers provided. Of these, four were UK 070 numbers, one was from Ohio, USA, and one from Nigeria. The UK 070 range is for 'personal numbers' allocated to people who continually change locations and can be answered pretty much anywhere in the world. These numbers are widely used by West African fraudsters because they appear to be UK numbers -- and our searching turned up previous reports of AFF scams associated with three of the four numbers.

\subsubsection{Prevalence}

In our UK sample, gathered over three weeks from Craigslist, 44 out of 2\,021 categorized letting advertisements (2.2\%) were rental scams for long-term letting that did not immediately redirect potential victims to a website. This number should be interpreted cautiously, because we only examined one classified advertising site in one country. In addition, we encountered several other scams, including holiday rental scams, letting scams using fraudulent websites, and advertisements where rent would be waived in return for sexual favours. The other scams were outside the scope of this paper and were not examined in detail. 

\subsubsection{What is on offer}

The majority of the 44 fake advertisements were for multi-room apartments (91\%), with the rest being single rooms (9\%). None of them were for entire houses, although houses are offered by genuine landlords on the same website. The apartments were located mainly (57\%) in London (25) with the remaining 19 being in Cambridge (5), Aberdeen (3), Liverpool (3), Manchester (2), Oxford (2), Bath (1), Brighton (1), Coventry (1) and Edinburgh (1).

On average, scammers offered their housing for \pounds 651.33 per month (SD = \pounds 332.78). The main goal of rental scams is making money, but because the scammers don't own the apartment, the victims will not end up living in it, and the scammers will not be able to charge rent except possibly for a month's rent in advance. Instead, when they ask victims to transfer money they give several reasons -- with most victims being asked to transfer money as a deposit to secure the apartment.

On average, the scammer mentioned the need for an upfront payment between their second and third email (M = 2.39, SD = .77, Range 1--5), and instructed the victim to make the transfer between the third and the fourth email (M = 3.52, SD = 1.18, Range~2--6). This could be a deposit plus one month's rent (32\%), just a deposit (23\%), two months rent (7\%), 2 months rent plus deposit (5\%) or the cost of posting keys (5\%); the remainder were unknown or not applicable.

On average, victims were asked to transfer a deposit of \pounds 1\,032.48 (SD = 625.26, Range \pounds 100--\pounds 2\,560) to secure the apartment, and they were asked to transfer this money in several different ways. Most popular was Western Union (21\%), followed by a redirect to a website or third party (11\%), MoneyGram (5\%), either Western Union or MoneyGram (5\%), a general bank transfer (5\%), or a mixture of methods (5\%); the remainder was unknown or not applicable.

In four cases the victim was asked to make a regular bank transfer to the scammer and in five cases they were asked to transfer via a website or other third party. The most popular request, made in 14 of the conversations (32\%), was for a transfer to be made not to the scammer, but to the victim herself or a trusted friend or family member, with the explanation that this demonstrates that funds were available. ID~47 gives a typical explanation of how to do this transfer via Western Union:	

\begin{quote}
ID~47. 
\textit{
1. Locate the nearest western union outlet around you.\\
2. Pick up the form of transfer and fill in the details which needs to be filled and return back to the post master.Details needed on the form are \{sender name;your name and address.\}\{receiver your friend's name and address\}\\
3. Return back to the post master with the cash you are transferring to your partner or friend and wait for the receipt to be returned back to you.\\
4. Scan the receipt to me so that i can forward it to my lawyer to check it if its truly genuine and available in your friend's custody.\\
5. After it has been confirmed valid, i will send you an email so that you can inform the receiver to go an pick up the money and thereafter, you'll schedule a convenient time for the viewing.}\\
\end{quote}

The point of this payment method is that the victim will be under the impression that their money can only be collected by the trusted person to whom they have wired it. However, with both Western Union and MoneyGram, the scammers can pick up this type of low-value transaction armed with only the details available on the scanned receipt, provided that they are somewhere in the appropriate country and the intended recipient has yet to collect it. In particular, there will be no need to display matching government-issue ID. People are seldom aware of how these money transfer systems work and fail to see how their transaction can go wrong. The rental scammer still needs an accomplice in the UK; but it is preferable to using a UK bank account that could well be closed if the police were to mount an investigation.

\subsubsection{Scam interactions}

Interacting with scammers provided an insight into the type of information scammers are interested in. We analysed this on three different levels: per email ($n=183$), per conversation ($n=44$), and per identified scammer ($n=21$).

In total, scammers asked the victim for more information in 159 out of 183 emails (85.9\%). The most frequently asked question was a proof of payment ability (42.1\%), for example by asking to transfer a deposit, followed by a method of contact (25.1\%), personal information (23.5\%), rental information such as move-in date (21.9\%), a confirmation that the victim wants to rent the apartment (20.2\%), reassurance that the victim will be a good tenant (16.9\%), a reference (6.6\%), return of the rental agreement (4.9\%), a proof of seriousness (3.8\%), a photograph (3.3\%), a suggestion of an alternative way to demonstrate payment ability (1.6\%), and when to arrange a viewing (1.1\%).

Although scammers asked questions in the majority of their emails (85.9\%), they typically only answered a subset of the victim's questions. The scammer answered at least one question in 59.0\% of all emails, and answered all outstanding questions in 39.3\%.

The scammer provided new or additional information himself in 154 out of 183 emails (84.2\%). For example, the scammer included an attachment in 19 out of 183 emails (10.4\%), such as apartment photos (13), a rental agreement (4), and the landlord's (i.e., the scammer's) photo (1) or a scan of the scammer's passport (1) -- no doubt the photo and passport came from other victims.

Scam conversations on average consisted of 4.91 emails (SD = 2.96; Range~1--13). As intended by our experimental design, most scammers stopped emailing after the third refusal of payment (38.6\%). The others gave up at other stages:

\begin{itemize}
\item 15.9\%	before the money request was made,
\item 11.4\%	after the first hesitation,
\item 13.6\%	after the second hesitation shown by the victim, and
\item 6.8\%	at some another stage.
\item 13.6\% of the scammers kept on emailing after the third refusal, with the result that we broke off the conversation.
\end{itemize}

On a conversation level ($n=44$), the pattern of questions is slightly different. The most frequently asked question was personal information (75\%), followed by proof of payment (71\%), reassurance that the victim will be a good tenant (68\%), rental information (68\%), method of contact (52\%), a confirmation that the victim wants to rent the apartment (45\%), reference (23\%), proof of seriousness (16\%), return of rental agreement (14\%), send a photograph (14\%), provide a suggestion of an alternative way to demonstrate payment ability (7\%), and when to arrange a viewing (5\%).

The difference between the email and conversation level occurs because while scammers ask many questions only once or twice, the request for a proof of payment ability was asked repeatedly. This is directly linked to the goal of making money, whilst the other questions merely serve to keep up appearances. 

To study the role of pre-existing scripts in rental scam conversations, we analysed email content on a scammer level ($n=21$). To test similarities, two or more conversations per scammer were needed. Out of 21 identified scammers, seven scammers fulfilled this criterion. For those, we compared emails from different conversations and measured similarities. We found that none of the seven scammers only sent unique emails. Instead, they used scripted emails during each conversation. Four scammers sent scripted emails without any adaptations, while five sent scripted emails in which they made small changes such as name and rent. However, not all emails were scripted: five out of seven scammers also wrote unique emails. The first two emails of six out of seven scammers were scripted with or without small adaptations. Typically, the scammer started the conversation and asked for the money transfer with scripted responses, but created unique emails once the conversation developed further and the victim showed hesitation about the requested transfer. This mirrors the sales pitch scripts taught to sales staff in legitimate volume sales businesses such as investment and home improvements. These sales staff are also taught a variety of techniques for rapport establishment, objection handling and closing, as described for example by Cialdini, and we believe that scammers are operating in much the same way.

\subsection{Social persuasion techniques}

We manually coded ten categories of social persuasion techniques for all emails that scammers sent, including Cialdini's six social persuasion techniques, Lea et al.'s `need and greed' and Stajano and Wilson's `distraction', `dishonesty' and `sympathy'~\cite{Stajano, Cialdini, Lea}. We analysed the presence of these techniques on each individual email ($n=183$). Then, because scam conversations may develop over time and different persuasion strategies may be used at different stages of the grooming processes, we also analysed the data on a conversation level ($n=44$). To measure whether persuasion techniques are scammer-dependent or whether similar techniques are being used across all scammers, we additionally analysed the persuasion data by scammer ($n=21$). 

The results are given in Table~\ref{tableone} and they clearly demonstrate that the social persuasion techniques established by Cialdini, that are used in marketing and other facets of life, are also frequently used in rental scams~\cite{Cialdini}.

\begin{table}
\caption{Prevalence of the persuasion techniques used in rental scams. Those labelled C are from Cialdini 1984, L from Lea et al. 2009 and S from Stajano and Wilson 2011.}
\label{tableone}
\begin{tabular}{clccc}
\\
	&			&	Per email		&	Per conversation	&	Per scammer \\
	&	Persuasion technique	&	($n=183$)	&	($n=44$)	&	($n=21$) \\
\hline
C	&	Commitment \& consistency &	53.5\%	&	81.8\% &	100.0\% \\
C	&	Liking			&	47.0\%	&	90.9\%	&	85.7\% \\
C	&	Authority		&	42.1\%	&	84.1\%	&	90.5\% \\
C	&	Scarcity		&	30.6\%	&	70.5\%	&	81.0\% \\
S	&	Sympathy		&	25.7\%	&	61.4\%	&	61.4\% \\
C	&	Reciprocity		&	12.6\%	&	43.2\%	&	47.6\% \\
C	&	Social proof	&	 7.7\%	&	25.0\%	&	38.1\% \\
L	&	Need and greed	&	 2.2\%	&	 9.1\%	&	14.3\% \\
S	&	Distraction		&	 1.6\%	&	 4.5\%	&	 4.8\% \\
S	&	Dishonesty		&	 0.0\%	&	 0.0\%	&	 0.0\% \\
\end{tabular}
\end{table}

However, three of the four persuasion techniques, `need and greed', `distraction', `dishonesty' which have been identified in other frauds were not used especially often in rental scams~\cite{Stajano, Lea}. The exception was `sympathy' (i.e., relying on people's innate tendency to be nice and willing to help). For example, scammers mentioned a death in the family or a previous bad treatment by (potential) tenants.

Our result is partly explained by looking at the scam as a whole rather than just the emails. Only 2\% of the emails contained an explicit appeal to the victim's need and greed by stressing how good a deal was on offer. However, the whole premise of the original advertisement was a nice apartment for a low price -- a direct appeal to need and greed.

To measure this we determined for each apartment if the price was significantly below market value or not. In 68\% of all fake advertisements, the rent was significantly below what can be expected for the region, making the apartment `a bargain'. 

The remaining two principles, distraction and dishonesty, were seldom used by rental scammers. Distraction was only used in 1.6\% of all emails, because the majority of emails were focused on letting-related topics. Distraction might be a more effective strategy when scamming someone face-to-face. As already noted, dishonesty is not relevant and was never used. 

\subsection{Two new scam categories}

\begin{figure}[!t]
\caption{ Social persuasion techniques (\%) analysed at the level of emails, conversations, and scammers.}
\centering
\includegraphics{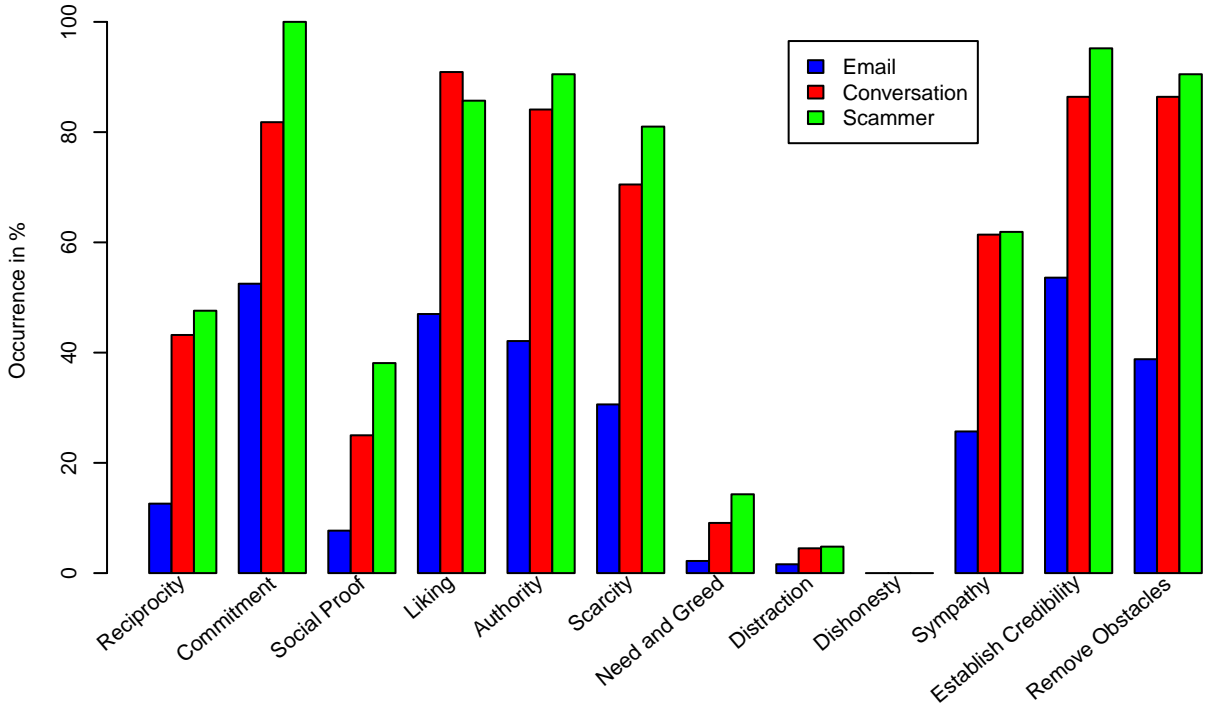}
\label{figure2}
\end{figure}

When coding the emails, we found that some types of persuasion did not fit in with the existing ten categories. This led us to add two new social persuasion categories: establishing credibility and removing objections. Both are familiar enough from everyday sales and marketing practice~\cite{Eades}. We present our complete analysis in Figure~\ref{figure2}, which displays the overall occurrence of social persuasion techniques in emails. 

\subsubsection{Establishing credibility}

From the deception literature we know that while truth tellers often take their credibility for granted, liars can feel the need to convince their interaction partner they are telling the truth~\cite{Kassin, Vrij}. Something similar seems to occur in rental scam communication, as we repeatedly found that scammers tried to establish their credibility. For example, a scammer would provide a plausible reason why he was not using a letting agent; he would offer to put the potential victim `in contact' with a satisfied previous tenant; he would provide another type of reference; he would pretend to arrange a viewing; he would share personal information and even passports, or would mention genuine organisations or initiatives such as the (real) Tenancy Deposit Protection Scheme (see ID~47). In some cases the scammer would even warn their victim about the risk of scammers and online fraud. This was all aimed at removing any doubt the potential victim might have about the legitimacy of the scammer.

\begin{quote}
ID~47. \textit{Rent per calendar month is \pounds 320 \{ includes all bills : water, gas, council tax, TV license and wireless Internet\} also Security deposit of \pounds 600 And it is protected by the Tenancy Deposit Protection Scheme (\url{http://www.depositprotection.com/}) \{Refundable after four weeks in the flat\}.}
\end{quote}

In our data set, establishing credibility was used in 53.6\% of all emails, making it the most prevalent technique at the email level. Furthermore, establishing credibility was used in 86\% of all conversations and by 95\% of all scammers. There is some overlap between Cialdini's `authority' and our `establishing credibility' because referring to an authority figure such as a lawyer or solicitor can help enhance credibility, but our new category contains enough original elements that we believe that it can be seen as a distinct persuasion category. 

Credibility establishment has previously been discussed in the marketing literature, and even in a scam context~\cite{Carter, Herbig}. Carter analysed the initial communication of letters and leaflets in postal scams, and identified that scammers promoted legitimacy and credibility in their communication. She also identified that scammers used `inspiring urgency', which corresponds to the way we have interpreted Cialdini's `scarcity', and also `secrecy', which is not applicable to rental scams.

\subsubsection{Remove objections} 

The second social persuasion technique we repeatedly came across was `removing objections'. In the real world, apartment and rental characteristics are often entirely inflexible; a flat is available from a specific date and it is furnished or unfurnished. Scammers do not actually have an apartment to rent out, so they can be flexible in what they offer. Through this removal of objections they can avoid potential victims deciding to drop out before the question of payment arises.

For example, the apartment offered by ID~66 was available both for short-term and long-term rental and you could stay in the flat as long as you paid your rent on time. There were further reassurances about the payment process, and a long list of every possible desirable item which would be provided free of charge, including a garage and parking spot in the middle of London -- and fresh towels on a daily basis.

\begin{quote}
ID~66. \textit{I can rent out the flat for unlimited time ( 1 to 5 years maybe more~). I can also rent it for short periods, the rent for 1 month is \pounds 950 per month including all utilities. You can use my furniture, or you can use your own, it's the same thing for me. I have the option of sending all my furniture to storage, if you want to bring your own (no extra cost). You can move in to the flat immediately.}
\end{quote}

In our data set, removing objections was used in 38.8\% of all emails, making it the fourth most prevalent technique at an email level. In addition, removing objections was used in 86\% of all conversations and by 91\% of all scammers. Both its prevalence and unique nature (removing objections does not overlap much with other, already existing social persuasion strategies), makes `removing objections' a useful addition to the scam principles found in the literature. 

We saw a few examples of negotiation and haggling, which are a way of removing objections when done from the scammer's (i.e., the seller's) side, rather than by the victim. For example, ID~58 offered his potential victim a deal: pay one month's rent plus deposit, or paying two months and getting an extra month for free. This is a classic `alternative question close'; ``would you like to pay cash for your new car, or apply for credit?'' Later, when the experimenter hesitated, the scammer dropped the demand for a deposit altogether, and offered to just send the keys for \pounds 100. So haggling, prevalent in sales negotiations, takes on an even more fluid form in the hands of scammers.
 
\subsection{Skilled scammers?}

The majority of previous scam communication research has been conducted on the scammer's initial email or letter~\cite{Onyebadi, Carter}. However, based on success rates and feedback, these initial emails may have evolved over time and may be used by large numbers of scammers. In other words, to send a successful first email, copy-paste skills may be more important than language and persuasion skills. By interacting with scammers in on-going email conversations, we have gained a unique insight into the scammers' skill-set. We will now assess how competent we believe the criminals to be; whether we felt they were sticking to a pre-determined script or whether they were capable of adapting to circumstances as the email conversation progressed.

There were a number of reasons for viewing rental scammers as skilful:

\begin{itemize}

\item
Scammers would regularly adapt to changing circumstances. For example, they answered at least one question from the victim in 82\% of all conversations, showing they paid attention to what the victim wanted to know. The scammers' responses also showed a high degree of flexibility and willingness to adapt their strategy. For example, upon the victim's hesitation about Western Union, scammer ID~77 provided general bank account details so payment could be made another way.

\item
Five of the seven scammers we held multiple conversations with constructed unique emails, allowing them to answer the victim's questions and remove objections.

\item
Rental scammers tried to mimic genuine house owners to enhance their credibility. For example, scammers provided new information in 84\% of their emails, and asked victims for information such as personal information, references and a method of contact in 86\%. Some scammers not only asked for a reference, but offered one as well. ID~25 went so far as to provide the email address of a `previous tenant'.

\item
Scammers mentioned the possibility of viewing in 70\% of all conversations, even though viewing would never take place. Viewing offers often came with an explanation why viewing was not immediately possible; the most common excuse was that the scammer moved away or lived elsewhere within the UK (41\%), followed by being abroad (36\%), that there were other people living in the apartment (7\%), and having had a previous negative experience (5\%). Only 9\% of scammers ignored the victim's viewing request altogether.

\item
Scammers generally used the anonymising email system provided by the Craigslist website, but they would also (in 55\% of all conversations) try to appear genuine by providing a specific method of contact, such as an email address, telephone number or Skype address. Six people provided a telephone number and four of those would appear to the uninitiated to be a UK number.

\item
Scammers have crafted a convincing method for victims to transfer money without asking victims to transfer money to them. Although some people might be cautious when Western Union or MoneyGram transfers are requested, they might not realise that the receipt contains sufficient information to pick up money that is sent in someone else's name. 

\item
Several scammers, including ID~92, highlighted the risks involved with money transfers and suggested using a third party to manage the deposit and key delivery (such third parties are of course part of the scam). Some scammers even referred to the rules of the `Landlord Association' to justify their request for an upfront payment.
\begin{quote}
ID~92. \textit{I am sorry but this is the only solution to this transaction. I must have a guarantee before I send the keys abroad. UPS is a serious company and they are known worldwide. They have a good reputation and your money will be 100 \% safe. I have used this service in the past and I was very satisfied. Please think about it and give me your final answer.}
\end{quote}

\item
The scammers made ample use of established social persuasions techniques; they tried to establish credibility and appeal to commitment and consistency in more than half of their emails. They also made an effort to be liked, appealed to authority, and removed potential objections in more than a third of all emails.

\end{itemize}

We conclude that rental scammers have some skill at their job -- and as noted large numbers of people fall for rental scams each year. On the other hand, there were also shortcomings in their persuasive techniques.

\begin{itemize}

\item
98\% of all scam conversations contained more than one spelling or grammar mistake, making it difficult to understand some of the emails and reducing their credibility.

\item
It was clear that scammers had often just copy-pasted a standard email, not just in the initial response but also further along in the conversation. So although scammers regularly (82\%) responded to at least one of the victim's questions across a conversation, they only answered all questions in 39\% of all emails. In other words, in two-thirds of the emails, at least one of the victim's questions was ignored. What seems to have been occurring is that the questions our `victims' posed related to the rental process and the pre-scripted emails `accidentally' answered some of these questions.

\item
Although the majority (71\%) of the seven scammers that we held more than one conversation with wrote some unique emails, none of them only sent unique emails. Instead, all scammers made use of scripted emails, and 57\% even sent scripted emails without making changes like adding the victim's name. This finding demonstrates that copy-pasting is an essential part of the scammer's toolkit. 

\item
Although scammers offered a viewing in 70\% of all conversations, these offers were only specific enough (i.e., proposing a day and a time) that an actual viewing would only be possible in 20\% of all conversations. Such specific offers were conditional on money changing hands first. Similarly, one can only view an apartment armed with the street address, and although in 89\% of all conversations the scammer mentioned some address details, they only provided the exact address in 18\% of conversations.

\item
Although some scammers asked victims to make a money transfer to a trusted friend, others asked for funds to be sent direct to them. Receiving a money transfer request before viewing will be considered a red flag by many potential victims, and several institutions and websites advise against make such a transfer~\cite{Craigslist, Trulia, Zillow, FTC}. That said, people who are not in the position to view an apartment may feel they have little choice.

\item
Scammers tended to be sloppy and made mistakes. For example, ID~73 sent the same email three times, ID~69 and 75 introduced herself with one name, and subsequently signed the email with another name, ID~22 mentioned conflicting information about the size of the rent, ID~111 addressed his second email to the wrong person, and another scammer sent the exact same email twice in a row, but once addressed to the wrong person.

\item
Scammers repeatedly used bogus arguments. For example, ID~24 claimed that a bank statement or employment contract was not an `acceptable' proof of payment ability in Westminster. Instead, the victim was asked to transfer money via Western Union. Similarly, ID~76 preferred Western Union transfers over regular bank transfers due to the ``high fraud occurrence in regular banks''.

\item
Although scammers used a wide variety of social persuasion techniques some were inappropriate for a British context. For example, scammers often appeal to authority by mentioning their lawyer or solicitor, and although this is a legitimate appeal to authority, it is not that relevant in the context of renting an apartment in the UK. Scammers also describe themselves as ``God fearing'' which may have been persuasive in Britain two generations ago but is not any more. In fact the word `God' has been identified as an effective filter-word against targeted scams on Craigslist~\cite{Park}.

\item
Although scammers asked for information in most of their emails, this was not the type of information genuine landlords request. For example, the scammers regularly wanted a proof of payment ability (42\%) at an early stage -- whereas this would normally occur much later in the process. Genuine house owners or letting agents would be more keen to determine the preferred move-in date. This disparity is of course understandable because the `proof of payment' will, as explained above, allow the scammer to walk away with the money as soon as possible.

\item
A scammer providing a reference might be seen a sign of scam expertise, as it can help to establish credibility. However, when looking more closely at the execution, one can wonder how convincing such a reference is if the language use and grammar/spelling mistakes are very similar to that of the scammer himself. 

\item
Sometimes the scammers' stories are not well thought out enough. For example, ID~92 suggested using UPS as a third party to ensure a safe transfer of keys and deposit, but when asked where information about this UPS service could be found, the scammer responded: ``I don't know where to find this service on their website either and I don't have a link to it'', reducing his credibility rather than enhancing it. Similarly, ID~84 offered to fly to the UK to show the victim around the apartment and sort out the contract. However, the flight the scammer said he booked did not exist. When confronted with this, the scammer stopped responding.

\item
Last, but far from least, scammers did not always provide enough information to make the money transfer. In a quarter of all conversations it remained unclear to the very end how exactly payment was to be made or how much money would be involved.

\end{itemize}

There is a striking parallel between the strengths and weaknesses of the scammers' persuasion techniques and those of legitimate businesses that rely on large numbers of sales staff to push their products. The authors have personal experience of industrial fundraising, and experience from student days of selling home improvements, insurance and advertising; we also have experience of listening to sales pitches from car dealers and others. The common pattern is that a sales organisation recruits staff who appear resourceful, personable and extravert; equips them with a sales script; trains them on objection handling; provides a number of techniques to close the sale; and incentivises particular business models. For example, when buying a car recently, one of us was struck not only by the similarity in sales pitches from staff at the dealers of different brands of car in different towns, but also that they all pushed hard to sell car finance and extended warranties, starting with the line that ``We are authorised by the Financial Conduct Authority, so we are required to ask you a few questions''. This behaviour is perfectly understandable as the commission from finance sales can exceed the margin on the vehicle itself. Similarly, the rental scammers' persistent attempts to get their deposit via Western Union rather than bank transfer is perfectly understandable in terms of the cost and inconvenience of replacing bank accounts that are closed following complaints of fraud.

Despite having multiple offenders targeting multiple victims in a systematic manner with a lot of common techniques the police appear to be dealing with this crime solely in a reactive mode on a case by case basis, as we will now discuss.

\subsection{Rental scam victims}

With the assistance of the University of Cambridge's Graduate Union (to which all graduate students belong), we invited victims of rental scams to fill out a questionnaire about their scam experience. We received four complete responses which help us understand the point of view of real victims and what happens when money is actually transferred.

All four victims are students at the University of Cambridge, between 18 and 25 years old. Two are male and two female. None of the victims was scammed more than once, and on average they lost \pounds 602.50. All four said they found out about the accommodation online, but were not referred to an additional payment website. Three out of four were scammed when distance-renting accommodation, and three out of four rented accommodation from someone who pretended to be the landlord.

The scammers had asked for the victims' name (100\%), their email address (100\%), mobile phone number (100\%), date of birth (25\%), current address (50\%), a scan of their passport (75\%) and their bank account details (50\%). All requested information was provided.

We were also interested to learn about the victims' decision process, and asked them about any suspicions they may have had. Two victims indicated that they had not been suspicious at all, one victim only got suspicious after transferring the money and one victim got suspicious near the end of the scam. 

We also asked about the financial and emotional impact of being scammed. On average they rated the financial impact as a 6.5 on a 10-point scale, but the emotional impact an 8.5. Rental scams can have serious negative impact on the victim's life, especially if the scam is only discovered when the new student is left standing in front of a closed door with all their belongings. This supports Button et al., who propose we should also take into account the emotional impact of scams~\cite{Button}.

After being scammed, all four victims contacted the scammer, the police and their bank. Half of the victims additionally contacted the website that hosted the advertisement, and a consumer protection agency. In three out of four cases, the police took a statement and the victims were told they would receive updates on progress. None of the victims received any updates -- but one eventually called the police back and was told some progress in solving the crime was being made. This dataset is too small to draw any general conclusions from but is in line with concerns about how effectively the British police tackle cybercrime, as identified by Her Majesty's Inspectorate of Constabulary in the Strategic Policing Requirement~\cite{HMIC}. The point is that individual frauds in the high hundreds to low thousands will generally be ignored, especially if there is an online or overseas element, as the police assume that investigation will be hard. However once these are seen as a pattern of organised crime, costing victims millions of pounds a year, the analysis and response may well be different.

\section{Discussion}
\label{sec4}

We have sought to provide an insight into the nature and prevalence of rental scams, and into the social persuasion techniques used during the scam process. At the start of this research, we were interested in whether rental scammers just follow a pre-determined script, or if they are skilled at the art of persuasion.

We scraped letting advertisements from Craigslist during a three-week period in early 2014, categorized each advertisement as `genuine', `fake' or `unsure', and followed up on the `fake' and `unsure' cases. We responded to scammers and pretended to be a potential victim. We gave the scammer three emails to convince us to transfer money to him.

We analysed 44 email interactions for scam characteristics and social persuasion techniques. We also asked four real-world rental scam victims about their experiences.

We found that scammers targeting the UK predominantly posted fraudulent adverts for property in London, and to a lesser extent in university cities such as Cambridge, Oxford, Liverpool, Manchester and Aberdeen. Victims are asked to pay a deposit or to transfer money to a friend by Western Union to prove they have the money -- which is then stolen using the information on the receipt. 

The sums involved ranged from \pounds 100 to \pounds 2\,560. Some scammers also directed payments to be made to third parties or to websites that they presumably controlled.

We identified ten social persuasion techniques in the literature and determined their prevalence at an email, conversation and scammer level. Scammers regularly implemented a wide variety of persuasion techniques, predominantly Cialdini's commitment and consistency, liking, authority and scarcity~\cite{Cialdini}. An appeal to sympathy, established by Stajano and Wilson was also frequently used to convince potential victims to transfer money~\cite{Stajano}.

Although Lea et al.'s appeal to need and greed was seldom explicitly present in the emails, it is the underlying assumption of most of the original fraudulent advertisements because the scammers are offering wonderful apartments at prices that are below the market rate~\cite{Lea}.

We also found that some of the scammers' persuasion techniques fell outside of the categories we found in the deception literature. We therefore added two techniques: `establish credibility' and `remove objections', both of which appear in the sales and marketing literature. The first of these relates to how scammers try to appear credible by providing explanations and references, referring to authority figures, and mimicking the actions of genuine landlords. The second describes the way scammers are far more flexible and accommodating at dealing with objections than any genuine landlord would be. Apartments turned out to be available on any date of our choosing, rental periods were entirely flexible and the space could be either `furnished or unfurnished'.

Establishing credibility as a persuasion technique has also been identified in postal scams and both how to appear credible and how to respond to a potential buyer's objections are key parts of training for legitimate sales and marketing activities~\cite{Carter}. Hence we believe that these two new categories are not specific to rental scams but will be found to be important types of persuasion in many other situations as well.

We assessed whether scammers are skilled at their job, or if they merely copy-pasted text that had worked in the past. Our repeated email interactions with rental scammers is, we believe, the first time that this has been studied.

Our results indicate that scammers operate very much like the sales staff at a legal business oriented towards interpersonal sales, such as car dealerships, insurance brokers and home improvements. Scammers start off their pitches using scripts that appeared to be widely shared among the cases we studied. They implement a wide variety of social persuasion techniques that have clear analogues in legitimate sales and marketing. They successfully impersonate genuine landlords by asking for rental-specific information, by answering the majority of the victim's questions, by offering the option of a viewing, and by providing a method of contact. They also use a clever way of getting hold of the victims' money, abusing a feature of money transfer systems that many potential victims fail to understand. The purpose of using this trick for obtaining the money is to minimise the cost to the scammers of replacing bank accounts that are blocked after victims' complaints. However, scammers could also be flexible. For example when we showed hesitation about a Western Union payment, the scammer provided general bank details for a regular transfer. 

However, when examining scammer behaviour in detail, it turns out their skills might be described as routine, and also limited by a lack of contextual and cultural knowledge (as might be expected if most of the scammers are operating from overseas, and merely have an agent in the UK to pick up the proceeds).

Even though they answered at least one question in the majority of emails, they failed to answer all questions in two-thirds. Even when the scammer sent a pre-scripted response, he often managed to `accidentally' answer a question. In some cases, even well into the email conversation, pre-scripted emails were still being sent. Although the scammer tried to mimic a genuine landlord, he did not always carry it out properly. For example, the scammer provided the exact address in only 18\% of all conversations, and still wanted advance payment before viewing. And although scammers did use a wide variety of social persuasion strategies, they failed to fit in with British culture. Their appeals to authority predominantly involved lawyers, solicitors and God in ways that were irrelevant to the way in which rentals usually proceed in the UK. Overall we found that though there was evidence of some skill, the scammers we interacted with provided enough warning signals to tip off a potential victim. 

Herley argued that online crime falls into two categories -- targeted attackers and scale attackers~\cite{Herley}. A typical targeted attacker is a highly skilled fraudster who sends phishing emails to a high-value target, such as the CFO of a company, with a view to taking over their computer and commanding unauthorised payments; such a fraudster may invest thousands of dollars' worth of time and effort, and if successful may manage to steal several hundred thousand dollars using wire transfers. A typical scale attacker is someone who infects large numbers of PCs using drive-by downloads and sells them to botnet herders; as the market price for compromised PCs is perhaps \$10 per 1000, the scale attacker cannot spend more than a nickel on each successful attack if he is to make money. Researchers had previously remarked on the apparent absence of attackers between these ``first class'' and ``economy class'' models of cybercrime.

Rental scams give an example of the missing middle. The skill levels required to conduct such scams are not zero, but neither are they astronomically high. Someone with experience as a street hustler or market-stall operator probably has the basics; add a working knowledge of written English and knowledge of crime scripts that have worked in the past, plus someone on the ground in the target country to collect the proceeds from a Western Union office, and a scam gang is in business. In short, setting up a 419 gang to do rental scams is not much different from setting up any other small selling business. This now appears to be an established modus operandi, and as noted the crime proceeds appear to be well into seven figures. Rather than considering it to be petty crime that is not worth more than a routine bookkeeping response, police forces should take it seriously and look for the bottlenecks where law enforcement pressure can be brought to bear -- whether on the organisers, the cash out operatives in target countries, or the money transfer firms themselves.

Payment service regulators in particular should consider imposing policy changes on companies such as Western Union and MoneyGram to require proof of identity on collection, which would increase the cost and risk to the scammer of operating in this way. 

\section{Conclusions}
\label{sec5}

At the operational level, this paper has documented how advance fee frauds are no longer implausible money-laundering schemes that only exceptionally gullible people would fall for, but have many new guises including rental scams that net millions a year. Although `419 scams', as they are also known, have been mostly targeted at the Western World, there is increasing evidence of activity in Asia and the Caribbean~\cite{Smith, Ultrascan, Akinladejo}. This is unsurprising because there are few barriers to targeting victims anywhere in the world other than possibly language.

The scientific work reported in this paper analyses the persuasion techniques that the scammers use. Previous work concentrated on the initial fraudulent communication whereas we have gone further by going through the whole scam process and giving the scammer three chances to ask for money before discontinuing the interaction. This has shown us the techniques scammers use, and they turn out to be not hugely different from people selling used cars, life insurance or home improvements. They have a mass prospecting technique, namely ads for apartments at competitive rentals in cities with a large transient population. They respond to inquiries from prospects using sales scripts, which are similar across multiple operatives. They use the standard sales and marketing techniques documented by Cialdini and other students of marketing, with a few twists that we describe. They have common techniques for dealing with objections. They have common sales closing strategies and techniques. Just as car salesmen push customers to buy on credit, so they can enjoy the finance sales commission as well as the margin on the vehicle, so also the scammers push people to use Western Union or MoneyGram, presumably to cut the costs of replacing blocked bank accounts. However they are flexible; just as a car dealer will accept cash if they have to, so also a scammer will provide a bank account to receive a deposit transfer if the mark stubbornly refuses to send money by Western Union. 

This is of independent interest to the cybercrime community as previously scammers had been argued to be either highly-skilled targeted attackers who go for high-value targets, or scale attackers who use fully automated techniques. The rental scammers give us a good example of criminals operating in the middle, using moderate levels of skill to defraud people of moderate amounts of money (high hundreds to low thousands) but on an industrial scale. These criminals show all the signs of being organised, just as legitimate businesses are. 

This has specific implications for policy. First, the police should treat rental scam complaints as evidence of a well-organised international crime, netting millions a year, rather than isolated fraud losses. Second, bank regulators should crack down on Western Union and other money transfer firms to insist upon total clarity about the risk of using these services. Third, we should explore ways to warn potential victims. In particular, universities should warn students about rental scams and inform them of accredited accommodation agencies.

\section*{Acknowledgments}

We thank the University of Cambridge's Graduate Union, Brett Ward and David Modic for their assistance in getting us in contact with rental scam victims. We also thank Francis Hsu for his assistance in setting up the email accounts we used.

The authors are funded by the EPSRC within the Deterrence of Deception project and by the Department of Homeland Security (DHS) Science and Technology Directorate, Cyber Security Division (DHSS\&T/CSD) Broad Agency Announcement 11.02, the Government of Australia and SPAWAR Systems Center Pacific via contract number N66001-13-C-0131. This paper represents the position of the authors and not that of any of the aforementioned agencies.

In accordance with the EPSRC's open data policy, an anonymised version of our rental scam data set will be made publicly available.

\bibliographystyle{abbrv}
\bibliography{GiftOfTheGab}

\begin{thebibliography}{10}

\bibitem{Akinladejo}
O.~H. Akinladejo.
\newblock Advance fee fraud: Trends and issues in the {Caribbean}.
\newblock {\em Journal of Financial Crime}, 14(3):320--339, 2007.

\bibitem{Buchanan}
J.~Buchanan and A.~J. Grant.
\newblock Investigating and prosecuting {Nigerian} fraud.
\newblock {\em United States Attorneys' Bulletin}, 49:39--47, 2001.

\bibitem{Button}
M.~Button, C.~Lewis, and J.~Tapley.
\newblock Not a victimless crime: The impact of fraud on individual victims and
  their families.
\newblock {\em Secur J}, 27(1):36--54, Feb 2014.

\bibitem{Carter}
E.~Carter.
\newblock The anatomy of written scam communications: An empirical analysis.
\newblock {\em Crime, Media, Culture}, 2015.

\bibitem{Chang}
J.~J. Chang.
\newblock An analysis of advance fee fraud on the {Internet}.
\newblock {\em Journal of Financial Crime}, 15(1):71--81, 2008.

\bibitem{Cialdini}
R.~B. Cialdini.
\newblock {\em Influence: The psychology of persuasion}.
\newblock William Morrow, New York, 1984.

\bibitem{Craigslist}
Craigslist.
\newblock Avoiding scams.
\newblock \url{https://www.craigslist.org/about/scams} (19 November 2019, last
  accessed).

\bibitem{Eades}
K.~M. Eades.
\newblock {\em The new solution selling}.
\newblock McGraw Hill, 2003.

\bibitem{Edelson}
E.~Edelson.
\newblock The 419 scam: Information warfare on the spam front and a proposal
  for local filtering.
\newblock {\em Computers \& Security}, 22:392--401, 2003.

\bibitem{FTC}
{Federal Trade Commission}.
\newblock Rental listing scams.
\newblock \url{https://www.consumer.ftc.gov/articles/0079-rental-listing-scams}
  (19 November 2019, last accessed).

\bibitem{Garrod}
S.~Garrod and M.~J. Pickering.
\newblock Why is conversation so easy?
\newblock {\em Trends in Cognitive Sciences}, 8(1):8--11, 2004.

\bibitem{Glickman}
H.~Glickman.
\newblock The {Nigerian} ``419'' advance fee scams: Prank or peril?
\newblock {\em Canadian Journal of African Studies / Revue Canadienne des
  ?tudes Africaines}, 39(3):460--489, 2005.

\bibitem{Herbig}
P.~Herbig and J.~Milewicz.
\newblock The relationship of reputation and credibility to brand success.
\newblock {\em Journal of Consumer Marketing}, 12(4):5--10, 1995.

\bibitem{Herley}
C.~Herley.
\newblock The plight of the targeted attacker in a world of scale.
\newblock {\em Workshop on the Economics of Information Security}, 2010.

\bibitem{HMIC}
{HMIC}.
\newblock Inspecting policing in the public interest. {T}he strategic policing
  requirement 2014, 2014.
\newblock
  \url{https://www.justiceinspectorates.gov.uk/hmic/wp-content/uploads/2014/04/an-inspection-of-the-arrangements-that-police-forces-have-in-place-to-meet-the-strategic-policing-requirement.pdf}
  (19 November 2019, last accessed).

\bibitem{Holt}
T.~J. Holt and D.~C. Graves.
\newblock Investigating and prosecuting {Nigerian} fraud.
\newblock {\em International Journal of Cyber Criminology}, 1:137--154, 2007.

\bibitem{Kassin}
S.~M. Kassin.
\newblock On the psychology of confessions: Does innocence put innocents at
  risk?
\newblock {\em American Psychologist}, 60(3):215--228, 2005.

\bibitem{Lea}
S.~Lea, P.~Fischer, and K.~Evans.
\newblock The psychology of scams: Provoking and committing errors of
  judgement. {Report for the Office of Fair Trading}, 2009.
\newblock \url{https://ore.exeter.ac.uk/repository/handle/10871/20958} (19
  November 2019, last accessed).

\bibitem{Modic}
D.~Modic and S.~E.~G. Lea.
\newblock How neurotic are scam victims, really? {The} big five and {Internet}
  scams, 2012.
\newblock \url{http://papers.ssrn.com/sol3/papers.cfm?abstract_id=2448130} (19
  November 2019, last accessed).

\bibitem{NFA}
{National Fraud Authority}.
\newblock Annual fraud indicator, {June 2013}, 2014.
\newblock
  \url{https://www.gov.uk/government/uploads/system/uploads/attachment_data/file/206552/nfa-annual-fraud-indicator-2013.pdf}
  (19 November 2019, last accessed).

\bibitem{Onyebadi}
U.~Onyebadi and J.~Park.
\newblock {`I'm Sister Maria. Please help me'}: A lexical study of 4-1-9
  international advance fee fraud email communications.
\newblock {\em International Communication Gazette}, 74(2):181--199, 2012.

\bibitem{Park}
Y.~Park, J.~Jones, D.~McCoy, E.~Shi, and M.~Jakobsson.
\newblock Scambaiter: Understanding targeted {Nigerian} scams on {Craigslist}.
\newblock In {\em 21st Annual Network \& Distributed System Security Symposium
  ({NDSS}), San Diego, CA, USA}, 2014.

\bibitem{NewsNigeria}
{PM News Nigeria}.
\newblock {Fred Ajudua} sent to jail, 11 June 2013.
\newblock
  \url{http://www.pmnewsnigeria.com/2013/06/11/fred-ajudua-sent-to-jail/} (19
  November 2019, last accessed).

\bibitem{Shelter}
Shelter.
\newblock 1m victims of landlord scams.
\newblock
  \url{http://england.shelter.org.uk/news_and_blogs/previous_years/2010/september_2010/1m_victims_of_landlord_scams}
  (19 November 2019, last accessed).

\bibitem{Smith}
R.~G. Smith, M.~N. Holmes, and P.~Kaufman.
\newblock Nigerian advance fee fraud.
\newblock {\em Trends \& Issues in Crime and Criminal Justice}, pages 1--6,
  July 1999.

\bibitem{Stajano}
F.~Stajano and P.~Wilson.
\newblock Understanding scam victims: Seven principles for systems security.
\newblock {\em Commun. ACM}, 54(3):70--75, Mar. 2011.

\bibitem{Independent}
{The Independent}.
\newblock The flat that never was: How the rental market became an attractive
  target for fraudsters, 12 November 2010.
\newblock
  \url{http://www.independent.co.uk/property/house-and-home/the-flat-that-never-was-how-the-rental-market-became-an-attractive-target-for-fraudsters-2131753.html}
  (19 November 2019, last accessed).

\bibitem{Trulia}
Trulia.
\newblock Rental listing scams -- read before you search.
\newblock
  \url{https://support.trulia.com/hc/en-us/articles/206731187-Rental-Listing-Scams-Read-Before-You-Search}
  (19 November 2019, last accessed).

\bibitem{Ultrascan}
{Ultrascan Advanced Global Investigations}.
\newblock 419 advance fee fraud statistics 2013, 2014.
\newblock
  \url{https://www.ultrascan-agi.com/assets/files/Pre-Release-419_Advance_Fee_Fraud_Statistics_2013-July-10-2014-NOT-FINAL-1.pdf}
  (19 November 2019, last accessed).

\bibitem{Vrij}
A.~Vrij.
\newblock {\em Detecting lies and deceit: Pitfalls and opportunities}.
\newblock Wiley, Chichester, England, 2008.

\bibitem{Whitty}
M.~T. Whitty.
\newblock The scammers persuasive techniques model: Development of a stage
  model to explain the online dating romance scam.
\newblock {\em British Journal of Criminology}, 53(4):665--684, 2013.

\bibitem{Zillow}
Zillow.
\newblock Beware of scams and other internet fraud.
\newblock
  \url{http://www.zillow.com/wikipages/Beware-of-Scams-and-Other-Internet-Fraud/}
  (19 November 2019, last accessed).

\bibitem{Zingerle}
A.~Zingerle and L.~Kronman.
\newblock Humiliating entertainment or social activism? {A}nalyzing scambaiting
  strategies against online advance fee fraud.
\newblock In {\em 2013 International Conference on Cyberworlds, Yokohama,
  Japan}, pages 352--355, 2013.

\end{thebibliography}

\end{document}